\documentclass{aa}
\usepackage{graphics}
\usepackage{epsfig}
\usepackage{natbib}
\bibpunct{(}{)}{;}{a}{}{,}

\begin{document}

\title{Evolutionary synthesis models of starbursts}
\subtitle{IV. Soft X-ray emission}

\author{M. Cervi\~no\inst{1} 
    \and J.M. Mas-Hesse\inst{1}
    \and D. Kunth\inst{2}}
\institute{Laboratorio de Astrof\'{\i}sica Espacial y F\'{\i}sica
    Fundamental (LAEFF-INTA), P.O. Box 50727, E-28080 Madrid,  Spain
    \and Institute d'Astrophysique de Paris, 98 bis Bd. Arago, F75014
    Paris, France}
\offprints{Miguel Cervi\~no}
\mail{mcs@laeff.esa.es}
\date{Received 12 April 2000, accepted -- }

\authorrunning{Cervi\~no, Mas-Hesse and Kunth}
\titlerunning{X-rays in starburst}

\abstract{ In this work we investigate the evolution of the X-ray emission
  of a cluster of single young massive stars with different metallicities.
  We have considered the X-ray contribution originated by the diffuse gas
  heated by the mechanical energy released by stellar winds and SN
  explosions as well as the X-ray contribution from SN remnants.  The
  resulting ionizing spectrum (i.e. $\lambda < 912$ \AA) has been used to
  compute the expected intensity of the nebular \ion{He}{ii} $\lambda$4686
  \AA. The observational ratio \ion{He}{ii}/H$\beta$ could be reproduced by
  the models assuming that a fraction of the mechanical energy produced by
  the star-formation episode is reprocessed by interaction with the ISM as
  soft X-ray radiation, contributing to the He ionization.  However, the
  discreteness of the stellar populations affects the ionizing flux and may
  be responsible for the observed dispersion of the ratio.  We have finally
  used the synthesis models to estimate the contribution of circumnuclear
  star-forming regions to the multiwavelength energy distribution in Active
  Galactic Nuclei, finding that the UV to soft X-ray continuum in many
  Seyfert 2 galaxies seems to be dominated by star-formation processes.
  } \maketitle

\section{Introduction}

In the last years convincing observational evidences have been collected
about the presence of starburst regions in or around active Seyfert~2
nuclei \citep[ and references therein]{Hecketal97,GDetal98}. It has been
found that most of their UV light originates in (circum-)nuclear
star-formation sites; the possible connection between the nuclear activity
and the properties of these starbursts is still a matter of debate.
According to the unified scheme of Seyfert galaxies, the active nucleus
should be hidden by an opaque torus in the case of Seyferts~2, explaining
why the collected UV light is dominated by young, massive stars. The low
degree of contamination by the active source in the UV provides detailed
information about the properties of the star-formation processes. Moreover,
extrapolating to the radio -- X-ray ranges the emission associated with the
starbursts, it should be possible to disentangle the fractional
contribution of both sources (the starburst and the active nucleus) at
different energy ranges.  Multiwavelength evolutionary synthesis models
normalized to the observed UV emission would be the ideal tool to perform
this analysis.

With these ideas in mind we started some years ago a program to extend our
evolutionary synthesis models (\citealp{AKS89}, Paper I, \citealp{MHK91},
Paper II, hereafter MHK, \citealp{CMH94}, Paper III, hereafter CMH) to
high-energy ranges (soft and hard X-rays, $\gamma$-rays). The predictions
of $\gamma$-ray emission were presented in \citet{Cervetal00}, and this
paper is devoted to the X-ray emission. It became evident from the
beginning that to properly reproduce the X-ray emission, both the
contribution of evolved binary systems and the heating of the diffuse
interstellar gas by the release of mechanical energy (by stellar winds and
supernova explosions), had to be considered.  Nevertheless, while High Mass
X-ray binaries contribute essentially to the hard X-ray range \cite[above
few keV,][]{vBetal99,VBV2000}, the soft X-ray emission is dominated by very
hot diffuse gas, heated by the release of mechanical energy from the
starburst \citep{Hecketal95, Hecketal96, Dicetal96, SS98a, SS98b, SS99,
  SSS01}.  We will therefore discuss in this paper the predicted soft X-ray
emission, and its effects on other observables, like the relative intensity
of Hydrogen and Helium emission lines. The contribution of binary systems
and evolved sources to the hard X-ray emission will be discussed in a
forthcoming paper.

Additionally, the statistical dispersion due to the discreteness of actual
stellar populations as presented in \citet{CLC00,CVGLMH01,Cetal01} is
particularly important in the high energy domain, where the number of
effective sources is lower. This computed statistical dispersion allows to
perform a better comparison with real systems and to evaluate the
statistically relevant sources in each energy range.

We present in Sec.~\ref{sec:model} our evolutionary synthesis model and
show how the X-ray emission and its associated statistical dispersion have
been computed. In Sec.~\ref{sec:output} we show the predictions on soft
X-ray emission. In Sec.~\ref{sec:lines} we explore the effects of the soft
X-ray contribution on the He~{\sc ii} nebular emission line.  In
Sec.~\ref{sec:agn} we compare our predictions with observational data from
star-forming and Seyfert galaxies, aiming to disentangle the relative
contribution of starbursts to the global energy budget of Seyferts. We
finally summarize in Sec.~\ref{sec:summ} our conclusions.

\section{The evolutionary synthesis model and the X-ray emission}
\label{sec:model}

We have updated and improved the synthesis model presented in
\citet{AKS89,MHK91,CMH94} with the following modifications:

\begin{enumerate}
\item Inclusion of the full set of Geneva evolutionary tracks including
  standard \citep{Schetal92} and enhanced mass-loss rates
  \citep{Meyetal94}.
\item Inclusion of metallicity dependent atmosphere models for normal stars
  from \cite{Kur}, CoStar \citep{SK97} and the atmosphere models for WR
  stars from \citet{Schmetal92}.
\item Inclusion of an analytical Initial Mass Function (IMF) formulation
  using a dynamical mass-bin\footnote{We have used a resolution in the HR
    diagram of $\Delta (\log T_{\mathrm{eff}})$ = 0.001 dex and $\Delta
    (\log L)$ = 0.001 dex for alive stars and $\Delta m$ not larger than
    0.1 and not lower than $2\times 10^{-7}$ M$_\odot$ for stars that have
    reached their life-time at the computed age.} \citep[see][ and
  references therein for a full description]{Cetal01}. We have also
  maintained the original Monte Carlo formulation.  
\item Use of parabolic interpolations in time for the computation of the
  isochrones \citep[see][ for a full study and discussion]{Cetal01}.
\item Evaluation of the dispersion due to the discreteness of the stellar
  populations in all the outputs (see \citealp{CVGLMH01} and
  \citealp{Buzz89} for more details).

\end{enumerate}

For this work, we will use only an analytical formulation of the IMF
without Monte Carlo simulations.  The basic input parameters of the model
are: {\it (i)} Instantaneous burst of star formation.  {\it (ii)} Salpeter
IMF ($\alpha=-2.35$) with a lower mass-limit equal to 2 M$_\odot$ and upper
mass-limit equal to 120 M$_\odot$ {\it (iii)} Set of evolutionary tracks
with standard mass-loss rates from \cite{Schetal92}.

In the following sections we describe how the different contributions to
the soft X-ray emission have been included in the computations.

We will consider as soft X-rays all photons with energies between 0.07 and
2.4 keV (the {\sc rosat} band), while hard X-ray photons will have energies
between 2.5 and 10 keV. We will also use the {\sc einstein} band (range 0.1
-- 3.4 keV) in Sec.~\ref{sec:agn} and the total X-ray luminosity (0.02--10
keV), $L_\mathrm{X}$.

\subsection{Sources of soft X-ray emission}

The main (persistent) contributors to the X-ray emission in a star-forming
region with only single-star populations will be essentially individual
stars, supernova remnants (SNR) and hot diffuse gas. In addition, other
sources like supernova explosions themselves, produce transient peaks of
X-ray emission. Since these peaks would be of very short duration, we have
not included them in our calculations, but we want to stress that they
could provide some degree of variability in the high-energy emission of
starbursts. Additionally, neither pulsars in young supernova remnants nor 
X-ray binaries have been included in the computations, as explained below.

\subsubsection{Individual stars}

In the case of massive stars, strong mass losses produce first a cloud of
material surrounding the star. In a second stage, winds from the star
shocks this material generating X-ray emission \citep{CG91}. The luminosity
in X-rays for stars hotter than B1 may be approximated by \citet{Cetal89}:

\begin{equation}
  L_\mathrm{X} = 10^{-7}~L_\mathrm{bol}
\end{equation}

The maximum contribution from the total population of individual stars is
about 10$^{30}$ erg s$^{-1}$ M$_\odot^{-1}$ at 3 Myr after the onset of a
burst, assuming an instantaneous burst following a Salpeter IMF slope. As
we will see, this contribution is around 2 orders of magnitude lower than
the emission from other sources and has not been taken into account for
computing of the total X-ray spectral energy distribution. We have instead
included the amount of mechanical energy released by the stellar winds in
the total budget of energy injected into the interstellar medium, as
discussed later.

\subsubsection{Supernova Remnants}

Supernova explosions (SN) will contribute strongly to the whole spectral
energy distribution, but only during a very short time (the light-curve of
a SN decays very quickly in few years).  This contribution is furthermore
very variable in a short time scale and has not been taken into account in
the models.

On the other hand, after a SN explosion, there will be a remnant of
expanding hot gas that will contribute essentially to the X-ray and radio
domains during its free-fall and adiabatic phases. The time scale of the
free-fall phase is about 90 yr \citep{Wol72}, and has not been taken into
account in the models. The adiabatic phase has a time scale of
\citep{Wol72}:

\begin{equation}
\tau \approx 2.0 \times 10^4 \epsilon_{50}^{4/17} n^{-9/17} {\mathrm yr}
\label{eq:tauSNR}
\end{equation}

\noindent where $n$ is the density where the SNR expands and
$\epsilon_{50}$ the mechanical energy released by the SN explosion in units
of 10$^{50}$ erg. Assuming a mean value of $\epsilon_{50}$ = 10 and $n=1$
atoms cm$^{-3}$, a SNR emits in X-rays during $3.4\times 10^4$ yr.
Following MHK, the contribution to the spectral energy distribution from
SNRs can be approximated by:

\begin{equation}
{\it L}^\mathrm{tot}_\nu \approx \dot{N}_\mathrm{SN} \int_{0}^{\tau}{\it
L}_\nu(t) dt = \dot{N}_\mathrm{SN} S_\nu
\label{eq:LSx}
\end{equation}

\noindent
where $\dot{N}_\mathrm{SN}$ is the SN rate, $\tau$ the time in which the
SNR shows X-ray (or radio) emission and $S_\nu$ the emission associated to
one individual SNR. Eq. \ref{eq:LSx} assumes that the supernova rate is
high enough to maintain a steady emission from SNRs during the burst.

It is important to point out that in the case of the analytical-IMF
computations, we have used a mass bin such that there is, at least, one
star suffering a SN explosion in every step of 10$^4$ years starting 2 Myr
before the computed age. This assures that the computation of the X-ray
emission is correctly made and that there is no bias due to the choice of
the mass bin.  The final output is the average value over the used time
step.

For the emission in radio and X-rays of SNR we have done the following
assumptions concerning $S_\nu$:

\begin{itemize}
  
\item{\it X-rays:} During the adiabatic phase of the SNR evolution, we have
  assumed it will emit soft X-rays following a composite Raymond-Smith
  hot-plasma law \citep{RS77} at different temperatures and at a
  metallicity consistent with the metal abundance of the region. In
  accordance with the compilation of \citet{Dicetal90}, we have assumed the
  same average temperature for the hot gas in the cavity of all SNR, as
  given by Eq.~\ref{ec:SN}:

 \begin{equation}
    \begin{tabular}{ll}
      $f(kT)_\nu^\mathrm{soft} =$ &
      $0.65 f(0.76 ~ keV)_\nu + $\\
      & $0.175 f(0.23 ~ keV)_\nu  +$\\
      & $0.175 f(1.29 ~ keV)_\nu$  \\
    \end{tabular}
    \label{ec:SN}
  \end{equation}
  
  where $f(kT)_\nu$ is a Raymond-Smith spectrum of temperature $kT$. We
  have normalized the emission of each individual SNR within the {\sc
    einstein} band (0.1--3.4 keV) to an average value of 10$^{36}$ erg
  s$^{-1}$, similar to the emissions given for the {\it Cygnus Loop} by
  \citet{Cox72} and \citet{CML90} for SNR in the Large Magellanic Cloud.
  With these values, and assuming that the emissivity within the {\sc
    einstein} band is roughly constant in the range of temperatures of the
  adiabatic phase, we obtain finally $S_{\mathrm{EINSTEIN}} = 3.4 \times
  10^{40}$ erg s$^{-1}$ year, which is the value we have considered in our
  calculations.
  
  We have assumed in addition a harder component with $kT = 6.4$ keV
  \citep{Cox72} originated at the shock front of the SNR, accounting for
  around 20\% of the total emission.
  
\item {\it Radio:} Following MHK, we have assumed that the radio emission
  originated by a SNR follows a power law with index $\alpha = -0.9$ from
  408~MHz to 4885~MHz (6 cm). We have updated the emission associated with
  a single SNR considered in MHK to a value at 6~cm of $3 \times 10^{9}$ Jy
  Kpc$^2$ year, that takes into account the possible contribution from
  cosmic rays and interaction with other remnants (see MHK and
  \citealp{MH92} for more details).

\end{itemize}

Individual point sources detected by {\sc rosat} in nearby galaxies are
indeed generally associated with SNR \citep{Ase98}.

\subsubsection{Mechanical energy released into the Interstellar Medium}

The large amounts of mechanical energy released by the starburst process
are expected to heat the interstellar medium around the newly formed stars.
Observations with {\sc einstein} already showed the presence of hot diffuse
gas within the H~{\sc ii} shells of 30~Dor \citep{WH91}. Combining {\sc
  rosat} and {\sc asca} data, \citet{W99} confirmed the thermal nature of
the emission, originated by diffuse gas at $2-9 \times 10^6$ K, in the soft
X-ray band. Similar results have been obtained in many other
starburst-dominated galaxies: \citet{SS98a,SS98b} found that the soft X-ray
spectra of a sample of 7 Wolf-Rayet galaxies could typically be well fitted
by a thermal Raymond-Smith component with temperatures in the range kT =
0.3--1.0 keV (corresponding to few times 10$^6$ K). These authors concluded
that a large fraction of the observed X-ray emission in these galaxies
originated in a hot superbubble formed by the combined action of stellar
winds from the massive, young stars in the central cluster.  \citet{SS99}
showed with hydrodynamical simulations that the observed soft X-ray
emission in NGC~5253 was consistent with the predictions for young
superbubbles blown by the starburst. Other starburst galaxies showing
thermal soft X-ray emission are: NGC~1569, with kT = 0.8~keV
\citep{Hecketal95}; Arp~220, kT = 0.61~keV \citep{Hecketal96} or
Henize~2-10 with kT = 0.31~keV \citep{Dicetal96} and Mrk~33 = Haro~2
\citep{SSS01}, among others \citep[see also][]{Hecketal97}.

We have computed the total release of mechanical energy from stellar winds
following the prescriptions in \citet{Leietal92}. The mechanical energy
released by SN explosions has been taken from the solar metallicity SN
simulations of \citet{WW95} for SN II and from the simulations of Helium
bare cores of \citet{WLW95} for WR stars. We have subtracted from the
energy of each SN explosion the total energy emitted in the 0.02 - 10 keV
band by the SNR during the adiabatic phase. It has a value in the {\sc
  einstein} band of $S_{\mathrm{EINSTEIN}} = 1.07\times10^{48}$ erg.

We have also assumed that a variable fraction $\epsilon_{\mathrm{eff}}^X$
of the total mechanical energy released will end up interacting with and
heating the diffuse interstellar gas to an average temperature of 0.5~keV.
As a first approximation we have thus considered a contribution to the
X-ray emission from starbursts coming from a Raymond-Smith thermal plasma
at a fixed temperature kT = 0.5~keV.  

Different estimations of $\epsilon_{\mathrm{eff}}^X$ can be found in the
literature based on hydrodynamical models, but they are quite dependent on
the assumed input physic \cite[see][ for an extensive review]{SS99,SS00}.
The standard bubble model of \citet{Weaver}, that assumes a constant energy
input into the bubble, shows that only 20\% of the energy injected into the
ISM is used in the bubble expansion (i.e.  $\epsilon_{\mathrm{eff}}^X \le
0.80$).  On the other hand \citet{SS99} find
$\epsilon_{\mathrm{eff}}^X$=0.05 from their hydrodynamical simulations
(assuming also a constant energy input into the expanding bubble).
  
We have chosen an arbitrary value of $\epsilon_{\mathrm{eff}}^X$=0.2 for
the presentation of this work. But we want to stress that the actual value
of $\epsilon_{\mathrm{eff}}^X$ should be derived by fitting the
observational values on an object by object basis. It is interesting to
note that a high value of $\epsilon_{\mathrm{eff}}^X$ leads to a high X-ray
emission, but, since the total energy in the burst must be conserved, it
also leads to a small radius for the expanding bubbles.  Therefore, the
comparison with observational data will only be consistent when both the
X-ray emission and the structure of expanding bubbles are taken into
account. The effect of changing the efficiency could be computed at any
time using the relative values given in Table~\ref{tab:cf}.

\subsubsection{Other X-ray sources}

\begin{itemize}
  
\item{\it High Mass X-ray Binaries (HMXRB)}: High Mass X-ray Binaries
  become X-ray ``active'' (permanent) after the primary star has collapsed
  in a compact companion, and when the secondary star enters the giant or
  supergiant phases, starting to accrete material onto the surface of the
  compact companion. HMXRB will become therefore active only after the
  first 4-5 Myr of evolution, depending on the upper mass limit of the IMF.
  The X-ray emission properties will be dependent on the radius of the
  compact companion.  Assuming a radius of the compact companion of around
  10 km, the bulk of the emission will be mainly in the hard X-ray band
  (2.5--10~keV) with a tail that will also emit in soft X-rays.  The hard
  X-ray predictions of our models including binary systems and their
  statistical relevance will be presented in a forthcoming paper.
  
\item{\it Pulsars in young SN remnants (YSNR)}: \citet{VBV2000} show that
  the loss of rotational energy of young pulsars in the form of
  electromagnetic waves are another possible source of X-ray emission. This
  emission can be fitted by a Raymond-Smith model with a temperature kT
  around 2~keV.  According to these authors, the total X-ray luminosities
  associated to these sources are around 10$^{32}$ erg s$^{-1}$
  M$_\odot^{-1}$ (for a starburst with a mass normalization given by an IMF
  with a slope $\alpha= -2.7$ and mass range 10 --100 M$_\odot$). This
  emission takes place only after the most massive stars have exploded as
  supernovae, and is negligible when compared with the total emission
  associated to both SNR and diffuse hot gas, even for rather low
  efficiency values. Furthermore, only a very small number of such objects
  is expected, so that their prediction is affected by very strong
  statistical fluctuations. For these reasons we have not considered this
  contribution in this work.
  
\item {\it Low-mass X-ray (LMXRB) binaries and Compton scattering}:
  \citet{PR02} compute the steady-state X-ray emission of star forming
  galaxies. In their work they include the emission produced by low-mass
  X-ray binaries, compton scattering and the emission of the compact
  nucleus in AGNs.  They also include the contributions from the diffuse
  emission and HMXRB.  Since our work aims to establish the contribution to
  the soft X-ray from star-forming regions, we have not included the
  emission of compact nuclei.  We are also restricted to young ages ($t <
  20$ Myr) and single stellar evolution, hence the contribution of LMXRB
  has not been considered. Finally, compton scattering is related to the
  electron density and the presence of magnetic fields and will emit mainly
  in the hard X-ray domain \cite[see][ and references therein]{PR02} and
  has not been included in the current computations.

\end{itemize}

As pointed out before, we have not considered these contributions,
but we keep in mind that the emission from YSNR could be significant in
starbursts devoid of interstellar gas, where no contribution from hot gas
is present and that HMXRB may increase the emission in both soft and hard
X-rays.

\subsubsection{X-ray conversion factors}

As far as the X-ray predictions are dependent on the assumed band, we have
summarized in Table~\ref{tab:cf} the conversion factors for the two
components discussed in this work normalized to the total X-ray luminosity
(0.02-10 keV).  Such conversion factors have been obtained by the
integration of the assumed X-ray spectrum over the given band. The results
for all the bands can be found in graphic and tabular form in our WWW
server at {\tt http://www.laeff.esa.es/users/mcs/SED}, but we will show
here only the predictions for the X-ray luminosity in the {\sc rosat} band,
for which most observational data are available. In the table we also give
the {\sc rosat} bands needed for the computation of the {\sc rosat}
hardness ratio.

\begin{table*}
\begin{tabular}{l|ccccc|c}
Metallicity & \multicolumn{5}{c}{\sc rosat}& {\sc einstein} \\
            & (0.07-0.4 keV) & (0.41-2.4 keV) & (0.41-1.0 keV) & (1.0-2.4
            keV) & (0.07-2.4 keV) & (0.1-3.4 keV) \\
\hline
 & \multicolumn{6}{c}{SNR} \\
0.001 & 0.290 & 0.441 & 0.251 & 0.190 & 0.731 & 0.745 \\
0.004 & 0.263 & 0.490 & 0.301 & 0.189 & 0.753 & 0.759 \\
0.008 & 0.243 & 0.526 & 0.337 & 0.188 & 0.769 & 0.770 \\
0.020 & 0.220 & 0.568 & 0.380 & 0.188 & 0.788 & 0.783 \\
0.040 & 0.207 & 0.591 & 0.403 & 0.188 & 0.798 & 0.789 \\
\hline
& \multicolumn{6}{c}{Diffuse hot gas} \\
0.001 & 0.408 & 0.457 & 0.360 & 0.097 & 0.865 & 0.813 \\
0.004 & 0.284 & 0.606 & 0.494 & 0.112 & 0.891 & 0.859 \\
0.008 & 0.221 & 0.683 & 0.563 & 0.120 & 0.904 & 0.883 \\
0.020 & 0.160 & 0.756 & 0.629 & 0.127 & 0.917 & 0.905 \\
0.004 & 0.134 & 0.788 & 0.658 & 0.130 & 0.922 & 0.915 \\
\end{tabular}
\caption[]{Values of X-ray luminosities in different bands normalized to
the total X-ray luminosity (0.02--10 keV)}
\label{tab:cf}
\end{table*}

\subsection{Influence on the ionization structure}

We have also taken into account that X-ray emission will produce a
relatively small fraction of ionizing photons (compared with those
generated by massive stars). Nevertheless, such photons will be
predominantly more energetic than the ones of stellar origin, 
and will affect the
ionization structure of the nebular gas, and thus the relative emission of
some lines, as we will discuss below. The number of ionizing photons
between the ionizing edges of H and He are given in Table~\ref{tab:qh} for
the two components considered, as obtained from the integration over the
assumed spectrum of each component.  In this case, we have normalized the
values to the total luminosity emitted in the {\sc rosat} band.  Note that
the normalization is different to the one used in Table \ref{tab:cf}.

As an example for any further use of this table, let us assume a Z=0.001
metallicity star forming region where the X-ray luminosity produced by SNR
is $5 \times 10^{38}$ erg s$^{-1}$ and the X-ray luminosity produced by the
diffuse hot gas is $5 \times 10^{39}$ erg s$^{-1}$, both in the {\sc rosat}
band. There will be an additional component in Q(H$^+$) of $2.07\times
10^{48}$ and $2.40 \times 10^{49}$ photons s$^{-1}$ due to the X-ray
spectrum from the SNR and the hot diffuse component respectively. Of
course, these additional Q(H$^+$) values are dependent on the assumed X-ray
spectrum, but it gives us a first order approximation of how important the
contributions of the X-rays sources to the ionizing flux are.

\begin{table}
\begin{tabular}{l|ccc}
Metallicity & Q(He$^{++}$)    &Q(He$^{+}$) &Q(H$^{+}$)  \\
            & ($\times 10^9$)& ($\times 10^9$)& ($\times 10^9$)\\
\hline
& \multicolumn{3}{c}{SNR} \\
0.001 & 2.35 & 3.80 & 4.13\\
0.004 & 2.24 & 4.05 & 4.38\\
0.008 & 2.17 & 4.24 & 4.36\\
0.020 & 2.11 & 4.43 & 4.74\\
0.040 & 2.05 & 4.53 & 4.85\\
\hline
& \multicolumn{3}{c}{Diffuse hot gas} \\
0.001 & 2.65 & 4.33 & 4.79 \\
0.004 & 2.07 & 3.50 & 3.87 \\
0.008 & 1.75 & 3.06 & 3.41 \\
0.020 & 1.49 & 2.69 & 2.99 \\
0.040 & 1.37 & 2.51 & 2.80\\
\end{tabular}
\caption[]{Values of the number of ionizing photons at different edges
normalized to the X-ray luminosity in the {\sc rosat} band (0.07 --2.4
keV), as originated by the two considered contributions (SNR and diffuse 
hot gas heated by the release of mechanical energy).}
\label{tab:qh}
\end{table}

\subsection{Evaluation of the dispersion}

In the following paragraph we summarize how to calculate the dispersion due
to the discreteness of the stellar population. We refer to \citet{Buzz89}
and \citet{CVGLMH01,Cetal01} for further details.

The IMF gives the probability, $w_i$, of finding a number of stars within a
given mass range. Each $w_i$ corresponds to the mass binning used in
analytical-IMF synthesis codes. If we assume that each $w_i$ follows a
Poissonian distribution \citep{CVGLMH01}, the variance, $\sigma_i^2$ of
each $w_i$ is equal to the mean value of the distribution, $w_i$. Let us
assume now that each star has a property $a_i$ whose contribution to the
integrated property $A$ is given by $w_i a_i$ with a variance $\sigma_i^2
a_i^2 = w_i a_i^2$. The total variance of the observable $A$ is the sum of
all the variances. The relative dispersion is:

\begin{equation}
\frac{\sigma_{A}}{A}=
\frac{(\sum_{i} w_i a_i^2)^{1/2}}{\sum_{i} w_i a_i} =
\frac{1}{\sqrt{N_{\mathrm{eff}}(A)}}
\label{eq:neff}
\end{equation}

\noindent where the last term gives us the definition of
$N_{\mathrm{eff}}(A)$ described by \citet{Buzz89}.  Note that
$N_{\mathrm{eff}}(A)$ is normalized to the total mass.
$N_{\mathrm{eff}}(A)$ is not a real number of stars, but is gives us an
idea of how many {\it effective} sources contribute to any given
observable. \citet{CVGLMH01} show that $N_{\mathrm{eff}}$ also defines the
mean value of a Poissonian distribution that can be used to obtain the
corresponding confidence levels of any observable in function of the amount
of mass transformed into stars.

Let us stress that IMF fluctuations are present in Nature (the number of
stars are always discrete), so, $N_{\mathrm{eff}}$ is not an evaluation
of the errors of the synthesis models. It is an evaluation of the
dispersion intrinsically present in real clusters, i.e. {\it the dispersion
  is also an observable}. This intrinsic dispersion must be taken into
account when fitting observed quantities to model outputs before
establishing any conclusion.

For differential quantities, like the SN rate or the mechanical
power, the obtained quantities are the mean value averaged over the 
time step (0.1 Myr in our case). The corresponding dispersion
shows the variation over such average mean value \citep{Cetal01}.

\begin{figure*}[htbp]
  \resizebox{\hsize}{!}{\includegraphics[width=18cm]{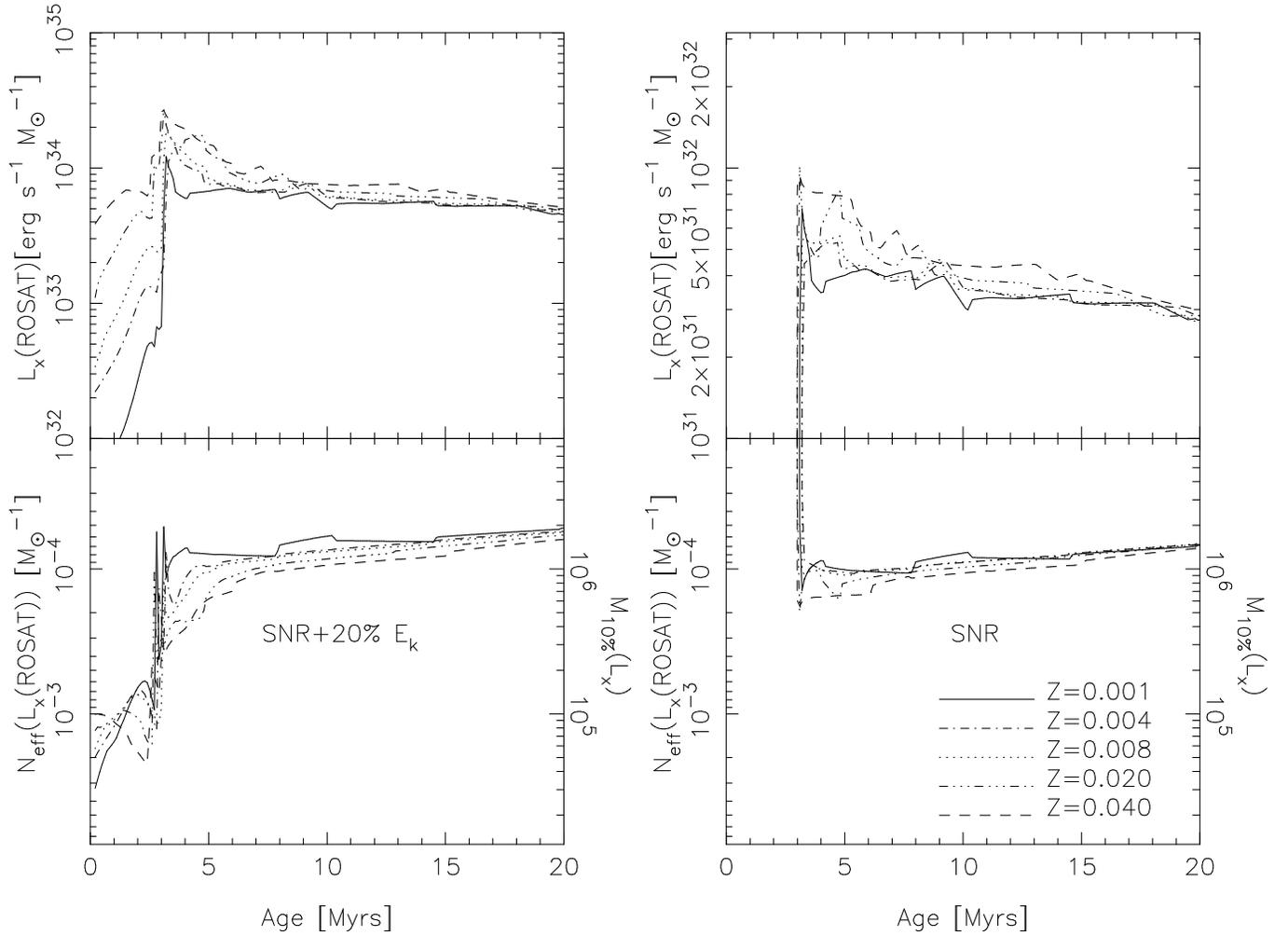}}
  \caption[]{{\bf Top panels:} X-ray emission in the {\sc rosat} band in
    erg s$^{-1}$ M$_\odot^{-1}$ as a function of metallicity for an
    instantaneous burst with Salpeter IMF slope in the mass range 2 -- 120
    M$_\odot$ for two cases: (a) SNR and $\epsilon_{\mathrm{eff}}^X$ equal
    20\%. (left) and (b) only SNR (right). {\bf Low panels}:
    $N_\mathrm{eff}$(L$_\mathrm{X}$) in units of M$_\odot^{-1}$. Right axis
    show the amount of gas transformed into stars needed to obtain a
    relative dispersion of 10\%, M$_{10\%}$(L$_\mathrm{X}$). }
\label{fig:X1}
\end{figure*}

\section{Predictions on soft X-ray emission}
\label{sec:output}

As discussed in the previous section, both supernova remnants and diffuse
hot gas will be the main contributors to the soft X-ray emission if only
single-stars are considered.  Depending on the evolutionary state of the
starburst, one or the other component will dominate the emission at a given
energy.  We show in the top panels of Fig. \ref{fig:X1} the predicted X-ray
emission in the {\sc rosat} band as a function of metallicity for two
cases: (a) X-ray emission produced by SNR plus hot gas, with
$\epsilon_{\mathrm{eff}}^X = 20\%$ in the left panel, and (b) the X-ray
emission produced by SNR only (i.e., with $\epsilon_{\mathrm{eff}}^X =
0\%$) in the right panel.

The X-ray emission depends on the star-forming region metallicity only
during the first few Myr, when the emission is dominated by the
reprocessing of mechanical energy, since stellar winds are strongly
metallicity dependent. On the other hand, after around 5~Myr the effect of
metallicity on the X-ray emission is small since the supernova rate is
essentially independent of metallicity.

It is clear that the total X-ray intensity is strongly dependent on the
associated $\epsilon_{\mathrm{eff}}^X$ value, as shown in the top panel of
Fig. \ref{fig:X1b}. It is remarkable that even a relatively low value
$\epsilon_{\mathrm{eff}}^X$ = 0.05 is enough to produce significant X-ray
emission, even during the first Myr of a starburst, when no other sources
are yet active. This is especially interesting considering that most
star-forming galaxies have been observed at ages between 3 and 7 Myr
\citep{MHK99}.  In general, a change in the reprocessing
$\epsilon_{\mathrm{eff}}^X$ from 0.05 to 1 implies approximately one order
of magnitude in total X-ray emission.

Since most starbursts seem to have formed massive stars according to an IMF
with a slope close to Salpeter's one \citep{MHK99}, and since their
metallicities can be derived from the analysis of optical emission lines,
it should be possible to derive a first order estimation of the average
$\epsilon_{\mathrm{eff}}^X$ value by just comparing the predictions with
the observed soft X-ray luminosities.

The {\it effective} mechanical energy and mechanical power remaining
available to drive gas flows away are smaller than the total amounts
generated by the starburst, since some fraction of the energy,
parameterized by $\epsilon_{\mathrm{eff}}^X$, is reprocessed into thermal
emission, and does not contribute to accelerate the gas.  If such effect is
omitted, the age deduced from kinematical studies will result
systematically lower than the one obtained from the global analysis of the
starburst.

\begin{figure*}
  \resizebox{\hsize}{!}{\includegraphics{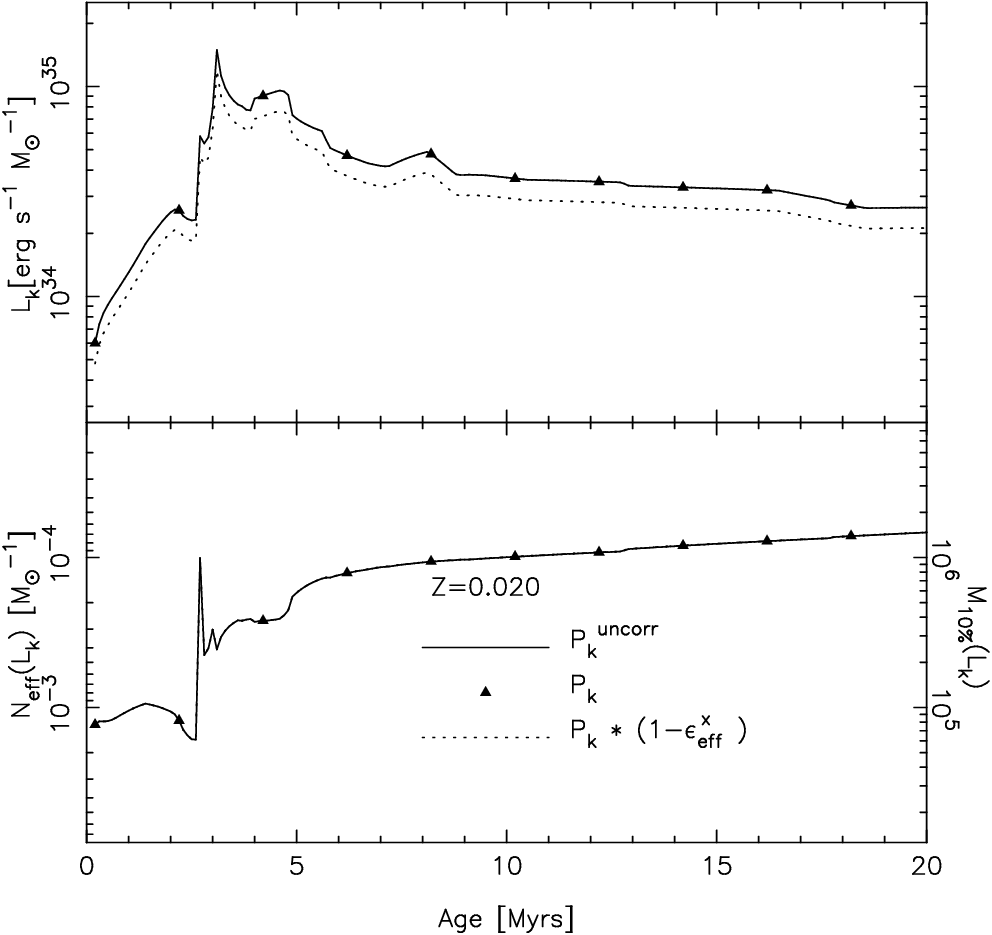}
    \includegraphics{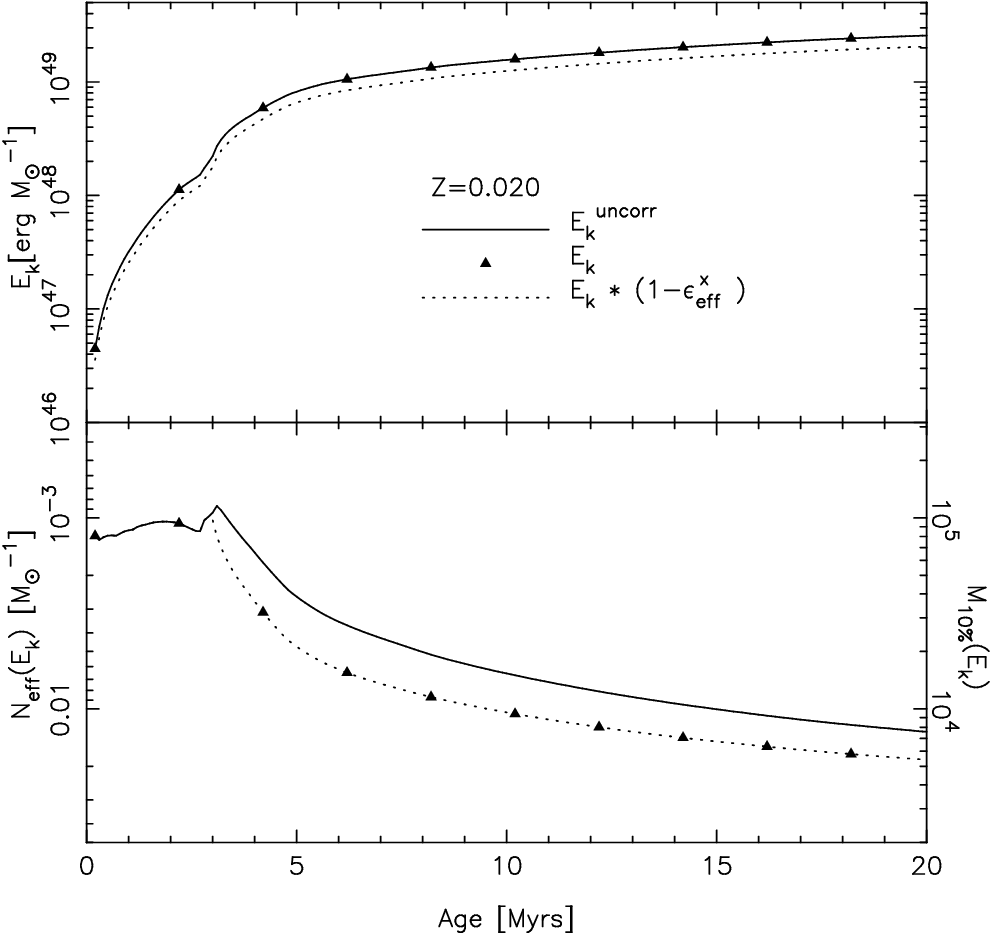} } \caption[]{Effective mechanical power and
    energy for solar metallicity and several hypothesis about the X-ray
    emission and its corresponding $N_\mathrm{eff}$(L$_\mathrm{K}$) and
    $N_\mathrm{eff}$(E$_\mathrm{K}$) in units of M$_\odot^{-1}$ and
    M$_{10\%}$(L$_\mathrm{K}$) and M$_{10\%}$(E$_\mathrm{K}$) values.}
  \label{fig:ek}
\end{figure*}

\begin{figure}
  \resizebox{\hsize}{!}{\includegraphics[width=18cm]{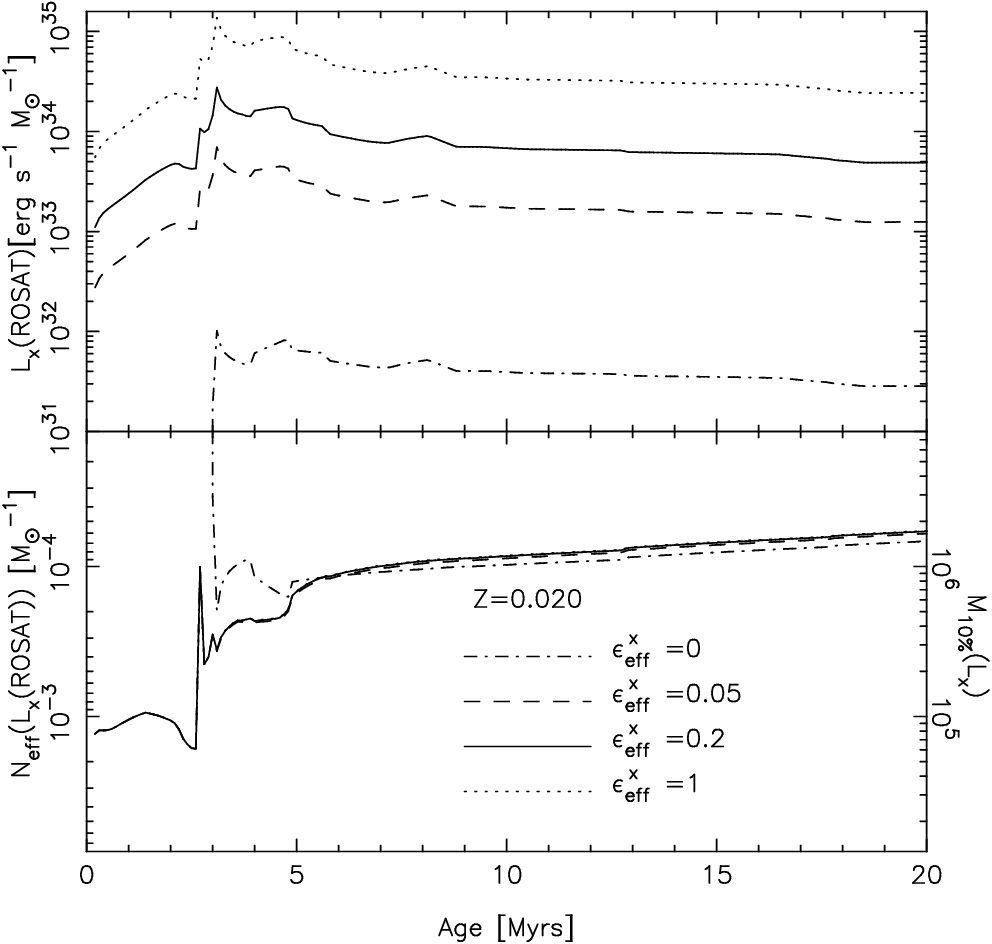}}
  \caption[]{X-ray emission in the {\sc rosat} band as a function of
    $\epsilon_{\mathrm{eff}}^X$, for an instantaneous burst with Salpeter
    IMF slope and solar metallicity and its corresponding
    $N_\mathrm{eff}$(L$_\mathrm{X}$) in units of M$_\odot^{-1}$ and
    M$_{10\%}$(L$_\mathrm{X}$) values.  }
\label{fig:X1b}
\end{figure}

In Fig.~\ref{fig:ek} we show the mechanical energy and mechanical power and
the corresponding {\it effective} mechanical energy and power when a value
of $\epsilon_{\mathrm{eff}}^X=0.2$ is used for the X-ray emission for a
solar metallicity burst. We also show these quantities when the correction
due to the X-ray emission of the SNR is not taken into account
(P$_\mathrm{K}^{\mathrm{uncorr}}$).

Lower panels of Figs.~\ref{fig:X1}, \ref{fig:X1b} and \ref{fig:ek} show the
value of $N_{\mathrm{eff}}(L_\mathrm{X})$, $N_{\mathrm{eff}}(L_\mathrm{K})$
and $N_{\mathrm{eff}}(E_\mathrm{K})$. The right axis shows the amount of
gas transformed into stars for the given IMF and mass limits needed to
ensure a dispersion lower than 10\%, denoted as M$_{10\%}$(L$_\mathrm{X}$),
M$_{10\%}$(L$_\mathrm{K}$) and M$_{10\%}$(E$_\mathrm{K}$). We will use this
notation for subsequent figures.

In the case of the X-ray luminosities and the mechanical power two regimes
can be separated. At the beginning of the burst the dispersion is dominated
by the stellar wind component, i.e. there is a large number of effective
sources that are contributing to the luminosity. When the first SN
explodes, the value of $N_{\mathrm{eff}}$ decreases abruptly: as far as the
remnant of a single SN produces more X-ray emission than the stellar winds
associated component, the effective number of sources decreases and the
dispersion due to the discreteness of the stellar population becomes
larger. It is also consistent with the fact that for evolved starburst the
possible dispersion will be dominated by the occurrence of SN events, and
so does the X-ray emission. In the case of the mechanical energy, the value
of $N_{\mathrm{eff}}(E_\mathrm{K})$ is larger (hence the dispersion lower)
because it is an integrated quantity \citep[see][ for more
details]{Cetal01}.

\subsection{Comparison with other works}

There are at least two other studies predicting the soft X-ray emission in
starburst galaxies: \cite{STTetal01} and \cite{SS99,SS00}. In both cases
the X-ray emission is obtained by simulations of superbubbles and
hydrodynamical models taking into account the time evolution of the kinetic
luminosity.

In \cite{STTetal01},  the X-ray luminosity is obtained from two
components: the interior of the superbubble and the shell, with
an analytical dependence of the metallicity on the radiative cooling
function, $\Lambda_\mathrm{X}(Z,T)$. They show that the soft X-ray emission 
depends on the enrichment of the hot ISM by stellar winds and SN
explosions. Nevertheless, their results are similar to the ones obtained in
\cite{SS99}.

The study by \cite{SS99} uses a superbubble model expanding in a vacuum
medium (i.e. an external pressure equal to zero) and obtains the X-ray
luminosity from the integration of the density structure and the radiative
cooling function over the volume of the bubble. They find a correlation
between the mechanical luminosity and the soft X-ray emission of
$L_\mathrm{X} (t) \approx 0.05 \times L_\mathrm{K}(t)$ with a time delay
between both quantities (variations in $L_\mathrm{K}(t)$ occurs earlier
than variations in $L_\mathrm{X}(t)$). They also present an extensive
discussion of the factors affecting the computed X-ray emission, assuming
superbubbles (see \cite{SS00}).  Their results (their Fig. 6) are quite
similar to the ones we have obtained (c.f. Fig.  2 scaled to a $10^6$
M$_\odot$ burst) with an efficiency of 5\%, except for the time delay as we
have explained above.

Additionally, 1-D hydrodynamical simulations computed by \cite{Plu01}
taking into account the environment where the bubble expands (an ambient
density of 40 atoms cm$^{-3}$ is assumed in their models), show that the
dissipation of $L_\mathrm{K}(t)$ is around 80\% (which includes not only
radiative cooling, but also the effects of mass-loading inside the
bubbles). Note that such effects may change the time delay between
$L_\mathrm{X}(t)$ and $L_\mathrm{K}(t)$.

In summary, our approximation parameterized in terms of $\epsilon^x_{eff}$,
although quite simplistic, provides a good first order
approximation to the more detailed superbubble simulations.

\subsection{The case of Mrk 33 = Haro 2}

As an example, we have applied our models to Mrk~33.  It is not the scope
of this paper to perform a detailed analysis of the properties of this
galaxy and it has been used only as a first order consistency check.  We
refer to \citet{SSS01} for a more complete study using different
constraints and synthesis models.

\begin{figure}
  \resizebox{\hsize}{!}{\includegraphics[angle=270]{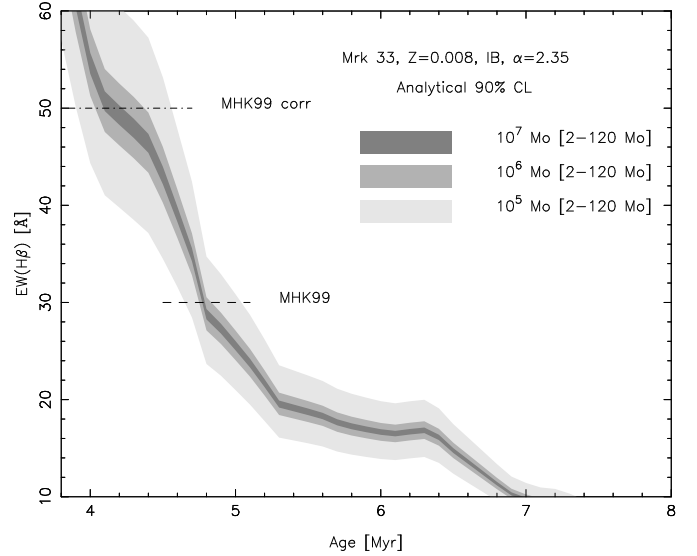}} 
  \caption[]{90\%
    confidence limits of W(H$\beta$) for different amounts of gas
    transformed into stars as a function of age, for a cluster with
    Z=0.008. We have overplotted the observed W(H$\beta$) value of Mrk~33,
    as given by \cite{MHK99}.}
\label{fig:mrk33hb}
\end{figure}

The first step to apply the model is to obtain an estimation of the age of
the burst. For this we refer to the W(H$\beta$) value used by \cite{MHK99},
which was obtained through a large aperture and was corrected from the
contamination by the underlying stellar population. We show in
Fig.~\ref{fig:mrk33hb} the predictions of the models as a function of age,
with the corresponding 90\% confidence limits for different amounts of gas
transformed into stars and a metallicity Z=0.008. The predicted W(H$\beta$)
values have been computed assuming that a fraction of 0.3 of ionizing
photons are not absorbed by the gas and the other 0.7 is transformed in
H$\beta$ luminosity following Case B recombination \cite[see][ for
details]{MHK99}.

As we can see in the figure, W(H$\beta$) alone constrains the age of the
starburst to around 4.2 Myr. By using a previous version of our set of
models, \cite{MHK99} derived an age around 4.8 Myr. For our case, it should
be enough to constrain the age of the starburst within the range 4 to 5
Myr, assuming an instantaneous burst.

The kinetic energy of the neutral expanding gas that is pushed out by the
shell in this galaxy is \cite[see][]{Leqetal95,Legetal97}
E$_\mathrm{k}$(H{\sc i})= $3 \times 10^{54} r^2$(kpc) erg, where $r$ is the
size of the expanding shell. Some possible values of $r$ have been proposed
in the literature: (i) \cite{Legetal97} derived an $r$ value of 1.23 kpc
from the expanding H$_\alpha$ shell and using a distance of 19.5 Mpc (ii)
\cite{SSS01} found a extended X-ray source with an ellipsoidal shape and
diameters of of 2.3 kpc and 1.9 kpc assuming a distance of 22 Mpc (i.e.
$r=\sqrt{ab}=1.0$kpc) and (iii). Also \cite{SSS01} obtain an average radius
1.1 kpc from B band images.  From this last study, we assume the same
distance and a value of 1.1 kpc and so a E$_\mathrm{k}$(H{\sc i})= $3.63
\times 10^{54}$ erg.
  
Finally, the observed soft X-ray luminosity in the {\sc rosat} band ranges
from 0.2 to 1.4 $\times 10^{40}$ erg s$^{-1}$. The first value corresponds
to the HRI observations of \citet{SSS01}, while the second one was derived
by \cite{SS99} from PSPC data.
  
Our model predicts E$_\mathrm{K}$ between $3.4 \times 10^{48}$ and
$5.4\times 10^{48}$ erg M$_\odot^{-1}$ at 4 and 5 Myr respectively, and
L$_\mathrm{K}$ between $7.0\times10^{34}$ (at 4 Myr) and $4.5\times
10^{34}$ erg s$^{-1}$ M$_\odot^{-1}$ (at 5 Myr). The uncertainty in
L$_\mathrm{K}$ is lower than 5\% in a 90\% confidence level and lower than
2\% (also in the 90\% confidence level) for E$_\mathrm{K}$ assuming a mass
transformed into stars, $M_{\mathrm{trans}}$, larger than 10$^6$ M$_\odot$
using a Salpeter IMF slope with mass limits from 2 to 120 M$_\odot$

A first order estimation of $\epsilon_{\mathrm{eff}}^X$ and the mass of the
cluster can be obtained using the following relations:

\begin{eqnarray}
    \epsilon_{\mathrm{eff}}^X &=& 1 -
    \frac{\mathrm{E}_\mathrm{K}^{\mathrm{obs}}}
    {\mathrm{E}_\mathrm{K}^{\mathrm{model}}\times M_{\mathrm{trans}}}
    \nonumber \\
    \epsilon_{\mathrm{eff}}^X &\approx & \frac{L_\mathrm{X}^{\mathrm{obs}}}
    {L_\mathrm{K}^{\mathrm{model}}\times M_{\mathrm{trans}}}
\end{eqnarray}
  
Note that, for a fixed value of $M_{\mathrm{trans}}$, larger values of
E$_\mathrm{k}$ produce larger values in $\epsilon_{\mathrm{eff}}^X$, and
lower values in L$_\mathrm{K}$ produce larger values in
$\epsilon_{\mathrm{eff}}^X$.

Assuming a constant $\epsilon_{\mathrm{eff}}^X$ value along the evolution,
it is possible to obtain $\epsilon_{\mathrm{eff}}^X$ values between 0.02
(at 4 Myr with HRI data and $M_{\mathrm{trans}}=1.1 \times 10^6$ M$_\odot$)
and 0.31 (at 5 Myr with PSPC data and $M_{\mathrm{trans}}=0.99 \times 10^6$
M$_\odot$). On the other hand, $M_{\mathrm{trans}}$ ranges from $0.7 \times
10^6$ M$_\odot$ (at 5 Myr with HRI data and
$\epsilon_{\mathrm{eff}}^X=0.04$) to $1.3 \times 10^6$ M$_\odot$ (at 4 Myr
with PSPC data and $\epsilon_{\mathrm{eff}}^X=0.15$).
  
So, the observed values can be explained with a 4--5 Myr old burst with a
mass transformed into stars of $0.7 - 1.2\times 10^6$ M$_\odot$ (following a
Salpeter IMF slope in the mass range 2 -- 120 M$_\odot$) and a
$\epsilon_{\mathrm{eff}}^X$ value from 0.02 to 0.31.
  
In this simple approximation we have not considered observational errors on
E$_\mathrm{k}$, neither different treatments of the mass-loss rates in the
evolutionary tracks nor different velocity laws in the computation of
E$_\mathrm{k}$ and L$_\mathrm{K}$, which would presumably increase the
range of possible values. So, a more careful analysis, based on a
self-consistent comparison of the {\it multiwavelength} spectral energy
distribution and a complete set of observables of Mrk 33 with the model
predictions is necessary to derive any firmer conclusion.  

\section{Predictions on optical emission lines intensities: H$\beta$ and
\ion{He}{ii}}
\label{sec:lines}

The H$\beta$ equivalent width has been used by many authors in the last
years as a reliable indicator of the evolutionary status of a starburst,
since it relates the most massive stellar population (main contributors to
the gas ionization) with the total population in the cluster, which
produces the optical continuum (see \citealp{Coppetal86} or \citealp{MHK99}
as an example).  Other emission lines have been proposed as age indicators,
like $[$\ion{O}{iii}$]$ ones by \citet{SL96}.  Additionally, the strong soft
X-ray flux produced by the hot diffuse gas and SNR will contribute to
harden the overall ionizing continuum, potentially affecting the ratios of
different emission lines.

\begin{figure*}
  \resizebox{\hsize}{!}{\includegraphics[width=18cm]{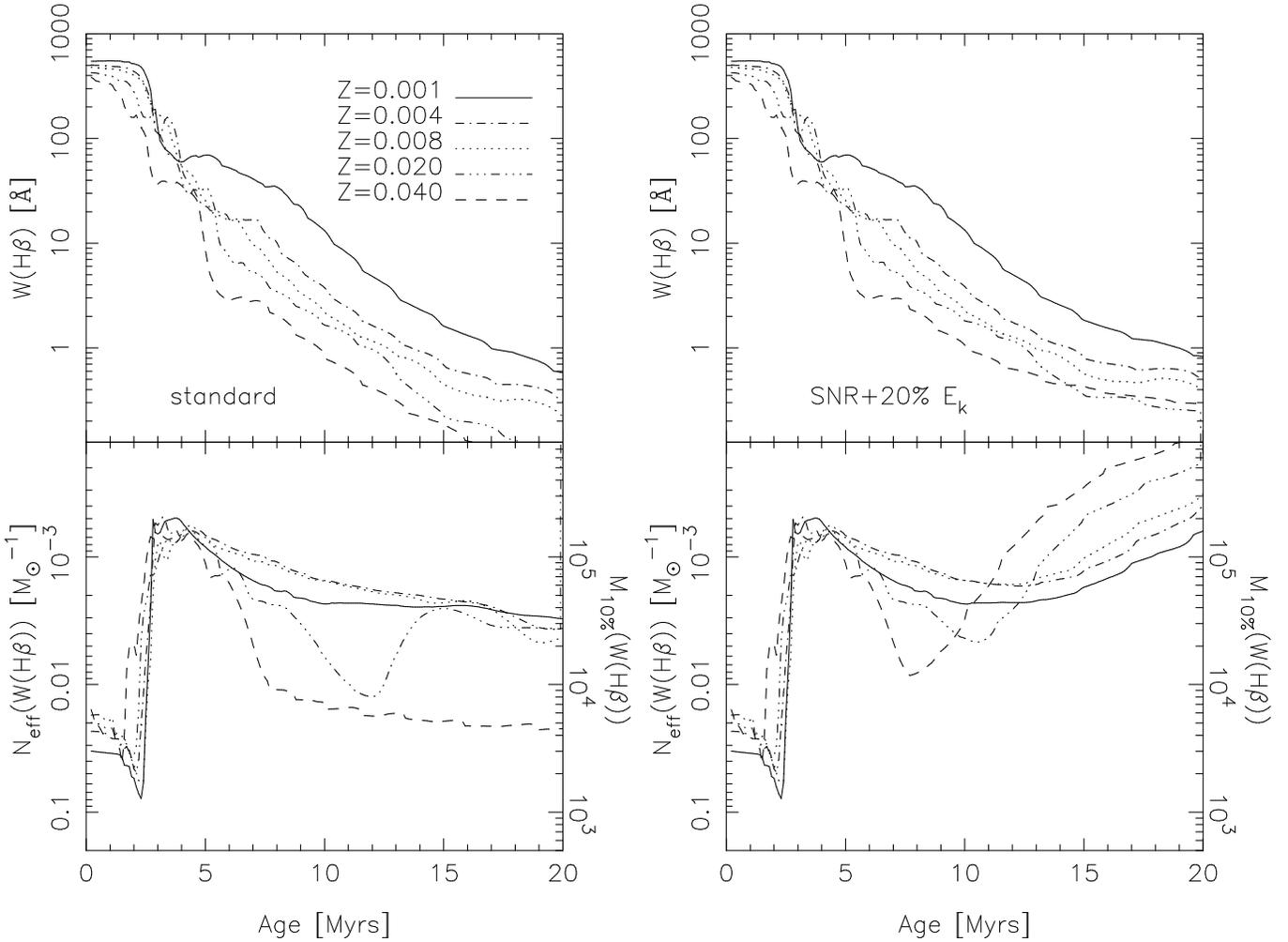}}
  \caption[]{W(H$\beta$) evolution as a function of the metallicity for a
    model where the X-ray emission is not taken into account and a model
    with a value of $\epsilon_{\mathrm{eff}}^X=0.2$ and its corresponding
    $N_\mathrm{eff}$(W(H$\beta$)) in units of M$_\odot^{-1}$ and
    M$_{10\%}$(W(H$\beta$)) values. }
 \label{fig:H}
\end{figure*}

We show in Fig. \ref{fig:H} the evolution with time of W(H$\beta$) as a
function of the metallicity for a model where X-ray emission (neither SNR
nor mechanical energy reprocessed in X-rays) is not taken into account
(standard) and a model with a value of $\epsilon_{\mathrm{eff}}^X=0.2$.  It
can be seen that the presence of hot gas may affect the predicted
W(H$\beta$) values after the first 6 Myr of evolution, i.e., when the
ionizing continuum of the massive stars decreases\footnote{The evaluation
of the differences in W(H$\beta$) between models that include X-ray
emission and standard ones depends on the metallicity and
$\epsilon_{\mathrm{eff}}^X$ values. We refer to the data in our WWW
server at {\tt http://www.laeff.esa.es/users/mcs/SED} for precise
comparisons.}.
A first implication of these results is that W(H$\beta$) becomes a very
uncertain age indicator for values below about 10~\AA.  A similar
conclusion has been obtained by \citet{vBetal99} based only on the effect
of binary systems, but in their case, the additional ionizing flux comes
both from the donor star of the binary system, that becomes (depending on
the mass-transfer scenario) a WR star, and from the gainer star, that
becomes more massive hence hotter than before the mass-transfer episode.
They obtain a higher value for a reliable use of W(H$\beta$) (i.e. not
dominated by binary systems) around 75~\AA.

Additionally, the dispersion of W(H$\beta$) when the X-ray emission is
taken into account becomes larger than the one without such hot gas
emission. It is due to the intrinsic uncertainty of the X-ray emission
itself (i.e. the small number of effective sources).  It also means that
the age determination of systems with low W(H$\beta$) values is an {\it
  intrinsically} difficult task, as far as the possible ionizing sources
are not only massive stars, but also diffuse hot gas.

While the inclusion of the soft X-ray flux in the ionizing budget does not
affect only weakly the total intensity of the H$\beta$ line, it can
affect more significantly other emission lines 
with higher ionization potentials, like
[\ion{O}{iii}] $\lambda$5007~\AA. At 54.9~eV, the soft X-ray flux might
become the dominant contribution to the ionizing continuum, affecting
significantly the expected intensity of the line.

The nebular \ion{He}{ii} 4686~\AA\ emission line has been detected in few
Star Forming Galaxies (SFG) \citep{Con91,SCP99}, but is apparently absent
in most of them.  The average ratio of the observed intensity of
\ion{He}{ii} over the intensity of H$\beta$, I(\ion{He}{ii})/I(H$\beta$),
is around 0.02 \citep{Fre80}.  Massive stars are not hot enough to produce
the hard ionizing continuum required for such relatively large ratios,
except perhaps during the WR phase, as discussed by \citet{SV98}.
Nevertheless, no clear correlation has been found between the detection of
the line and the presence of WR stars in the region.  Additional, harder
contributions to the ionizing continuum are therefore required to explain
the observed ratios.

\begin{figure*}
  \resizebox{\hsize}{!}{\includegraphics{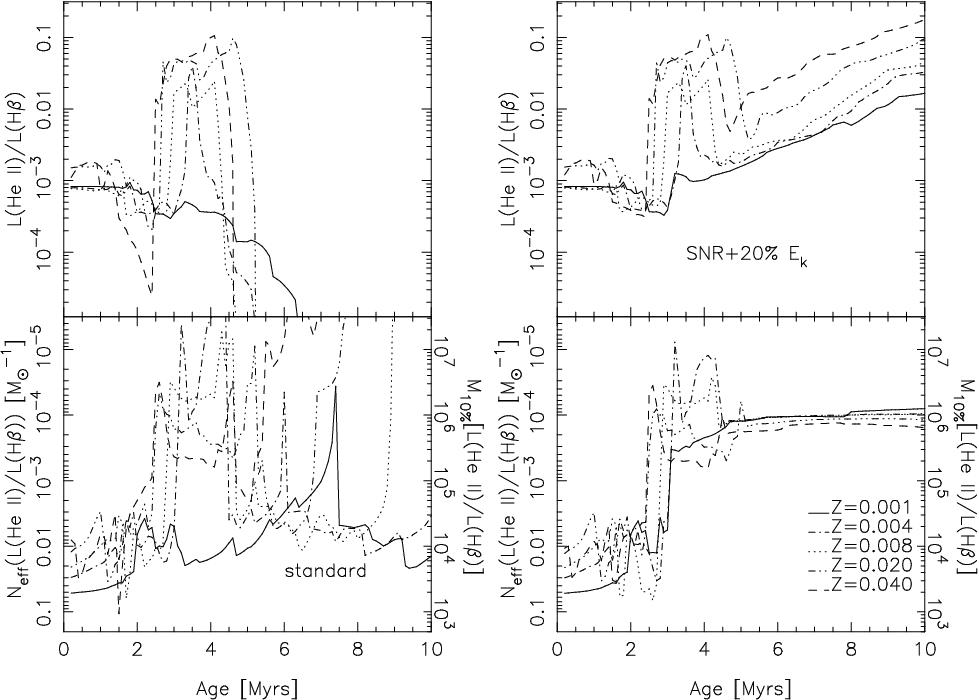}}
  \caption[]{I(\ion{He}{ii})/I(H$\beta$) ratio as a function of age and
    metallicity for a model where the X-ray emission is not taken into
    account and a model with a value of $\epsilon_{\mathrm{eff}}^X=0.2$ and
    its corresponding $N_\mathrm{eff}$(I(\ion{He}{ii})/I(H$\beta$)) in
    units of M$_\odot^{-1}$ and M$_{10\%}$(I(\ion{He}{ii})/I(H$\beta$))
    values.}
  \label{fig:He}
\end{figure*}

We have explored the effects associated with the reprocessing of mechanical
energy in the interstellar medium and the X-ray emission of SNR. The
results are shown in Fig.~\ref{fig:He}.

The ionization by hard photons produced by the diffuse gas heated by the
release of mechanical energy into the medium can lead to significantly
higher values of the I(\ion{He}{ii})/I(H$\beta$) ratio specially for low
metallicity clusters. As has been shown by \citet{SV98}, for solar
metallicity starburst, a value of the ratio of 0.01 can be explained only
with the presence of WR stars, but such value is never reached in low
metallicity clusters. We show here that it is possible to obtain such ratio
if the soft X-ray emission is included.

However there are some points we would like to note here:

\begin{itemize}
  
\item The extension of the ionized region emitting the \ion{He}{ii}
  4686~\AA\ line is generally much smaller than the region over which
  H$\beta$ is emitted, as shown by \citet{MAetal98} in NGC~4214. To
  properly compare with the predictions of synthesis models, the observed
  \ion{He}{ii} 4686~\AA\ flux has to be divided by the H$\beta$ intensity
  integrated over the whole ionized area. Such integrated ratios can be
  smaller by almost an order of magnitude than the ratios measured just
  locally.  Another illustrative example can be found in \citet{LPL99}
  where the authors apply a photoionization model to NGC 2363 with the aim
  of explain with a single model {\it all} the observations of the region
  obtained by other authors using different apertures. Whereas the observed
  I(\ion{He}{ii})/I(H$\beta$) ratios varies from 0.010 to 0.004, the global
  I(\ion{He}{ii})/I(H$\beta$) ratio of the model that fit all the
  observations has a value of 0.001.
  
\item Our models compute the predicted soft X-ray emission assuming the
  released mechanical energy is reprocessed immediately into high energy
  photons. A more realistic scenario would imply that the mechanical energy
  is first accumulated in the form of energetic gas outflows, being
  released in the form of soft X-rays only when this expanding gas
  interacts with static interstellar clouds. Therefore, at some stages, the
  soft X-ray contributions to the ionizing continuum could be stronger than
  values predicted by our models.
  
\item The hard ionizing flux intrinsically has a big dispersion. Its value
  is controlled by a small number of effective sources (WR stars, hot gas
  clouds or binary systems), hence suffers from severe statistic problems.
\end{itemize}

\begin{figure}
  \resizebox{\hsize}{!}{\includegraphics{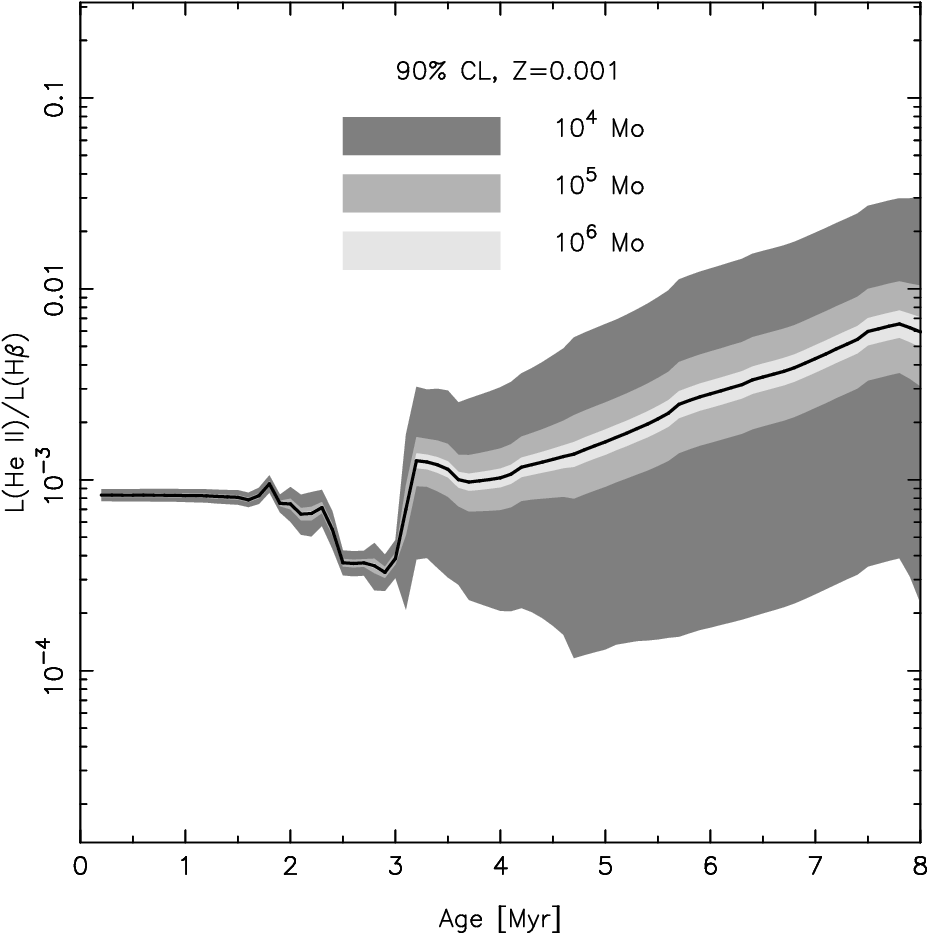}} \caption[]{90\%
    confidence levels of the I(\ion{He}{ii})/I(H$\beta$) ratio as a
    function of age for a Z=0.001 model with a value of
    $\epsilon_{\mathrm{eff}}^X=0.2$ for a cluster where 10$^4$, 10$^5$ and
    10$^6$ M$_\odot$ have been transformed into stars. We also show the mean
    value.}  \label{fig:HeCL}
\end{figure}

In order to support the last items, we show in Fig.~\ref{fig:HeCL} the
90\% confidence limits obtained from the
$N_\mathrm{eff}$(I(\ion{He}{ii})/I(H$\beta$)) value for a cluster with a
metallicity Z=0.001 and different amount of gas transformed into stars.

We therefore conclude that the additional hard ionizing photons produced by
the diffuse gas can explain, at least partially, the intensities of
\ion{He}{ii} 4686~\AA\ observed in low metallicity star forming regions.
However, at this point it is impossible to establish the source(s) of the
\ion{He}{ii} 4686~\AA . From a theoretical point of view, it is necessary
to include other sources of X-ray emission that were not considered in this
work, and establish their {\it statistical relevance}.  From the
observational point of view, more observations in different wavelengths
domains are needed to constrain the sampling effects due to the
discreteness of the stellar population. They must be unaffected by
additional aperture effect biases.

\section{Star-formation processes in Seyfert galaxies}
\label{sec:agn}

\begin{figure}
  \resizebox{\hsize}{!}{\includegraphics[angle=270]{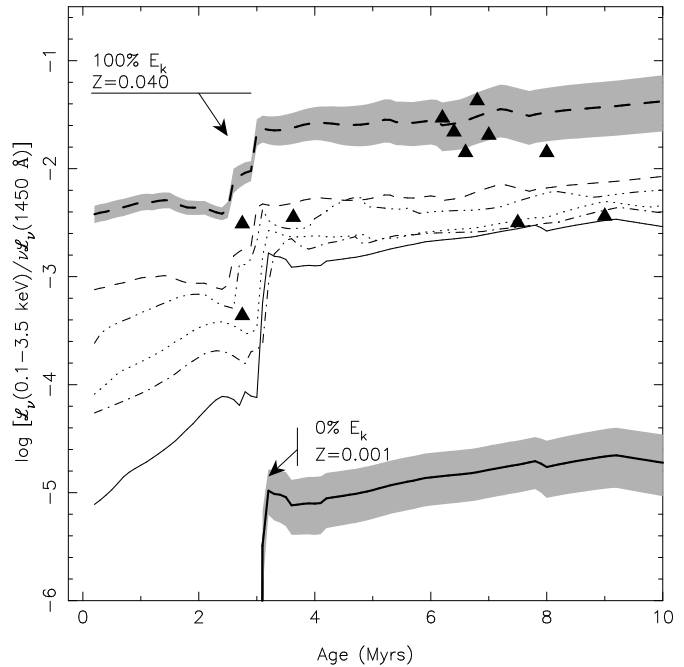}}
\caption[]{Ratio L$_\nu$(0.1 -- 3.5 keV) over $\nu$L$_\nu$(1450 \AA )
  computed for an efficiency in the reprocessing of mechanical energy into
  soft X-rays emission of $\epsilon_{\mathrm{eff}}^X$=0.2 and different
  metallicities (with lines as in Fig.~\ref{fig:He}). In grey, the
  90\% CL for a cluster where $10^5$ M$_\odot$ have been transformed into
  stars for two extreme metallicities and $\epsilon_{\mathrm{eff}}^X$
  values.  Data points are from SFG in \citet{MHetal95} compilation.  Note
  that the UV continuum of the data points has not been corrected from
  internal extinction.}
\label{fig:gal}
\end{figure}

We have compared the predictions of our models with the multiwavelength
data for a sample of AGNs (QSO, Seyfert 1 and Seyfert 2) and SFGs compiled
by \citet{MHetal95}.  We show in Fig.~\ref{fig:gal} the predicted evolution
of the {L$_\nu$}(0.1 -- 3.5 keV) over {$\nu$L$_\nu$}(1450~\AA) ratio for
$\epsilon_{\mathrm{eff}}^X$=0.2 and different metallicities. The model
predictions correspond to a Salpeter IMF.  We have also plotted the highest
(0.040 metallicity with $\epsilon_{\mathrm{eff}}^X$=1) and the lowest
(0.001 metallicity with $\epsilon_{\mathrm{eff}}^X$=0 including the SNR
component only) values of the ratio, and the corresponding
90\%CL for a 10$^5$ M$_\odot$ cluster, so that, for a given age, both solid
lines (and bands) represent the upper and lower limits we should expect
associated to a star-formation episode.  Data points are from SFG in
\citet{MHetal95} compilation.  We have determined the ages of the plotted
objects from the W(H$\beta$) values taken from the literature, but it can
be seen that in any case the results are weakly dependent on the assumed
age. UV data have not been corrected for internal extinction, so that the
plotted ratios have to be taken as upper values and should be indeed
intrinsically lower by factors between 2 and 5, approximately
\citep{MHK99}.

From Fig.~\ref{fig:gal} we see that almost all galaxies fall within the
predictions of synthesis models.  It is important to stress that {\em some}
conversion of mechanical energy into X-rays is always required.  Otherwise
the models would severely underestimate the observed X-ray/UV ratios.
Additionally, some galaxies fall outside the model limits. It can be due to
statistical effects and/or the presence of binary systems that will
increase the X-ray emission and/or the underestimation of the observed UV
flux due to extinction effects.  We want to point out that the effect of
the extinction in our sample has not been considered. Even a moderate
extinction would significantly affect the UV continuum, putting so the
X-ray/UV ratio within the model limits without additional X-ray sources.

Note that our models compute only the persistent X-ray emission form
single-stellar populations. Transient sources like Be/X-ray binaries and SN
explosions may increase the X-ray emission in these galaxies, but only
during very short periods of time. In the case of Be/X-ray binaries the
component may be highly variable and would affect only the soft X-ray
emission.

\begin{figure}
\begin{center}
  \resizebox{!}{!}{\includegraphics[width=5cm]{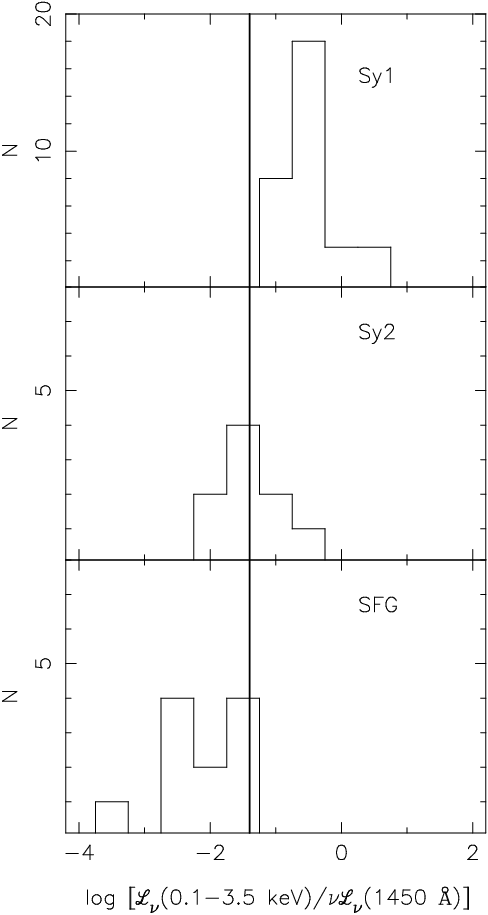}}
\end{center}
\caption[]{Histograms for the ratio L$_\nu$(0.1 -- 3.5 keV) over
  $\nu$L$_\nu$(1450 \AA ) for different types of emission line galaxies.
  Data taken from \citet{MHetal95}. The UV continuum of the galaxy sample
  has not been corrected from internal extinction. The solid line
  corresponds to the upper value predicted by our evolutionary synthesis
  models, as explained in the text. }
\label{fig:histo}
\end{figure}

In order to disentangle the relative contribution of (circum-)nuclear
star-formation processes to the total energy budget of low activity AGNs,
we have compared their L$_\nu$(0.1 -- 3.5 keV) over $\nu$L$_\nu$(1450 \AA )
ratios with the predictions of our synthesis models. We show in Fig.
\ref{fig:histo} the corresponding histograms for the Sey~1, Sey~2 and SFGs
taken from the \citet{MHetal95} sample. The vertical line corresponds to
the maximum value predicted by the models, as shown in Fig.~\ref{fig:gal}.

The segregation of Sey~1 and star-forming galaxies becomes evident at first
glance: while most of the SFGs in the sample fall below the predicted upper
value, all Seyfert~1 galaxies show clearly higher values, by one to two
orders of magnitude. Therefore, while the physical processes assumed to
take place in star-forming galaxies may explain their observational
properties, other high energy sources are required in the case of
Seyferts~1, as expected.

The case for Seyfert~2 galaxies is more interesting: while some of them are
located on the plot in the region associated to starbursts, some others
show a clear X-ray excess with respect to star-formation dominated objects.
The presence of star-formation regions in or around the nuclei of some
Seyfert~2 galaxies has become evident in the last years. \citet{Hecketal97}
showed that around 70\% of the UV continuum in some Seyfert~2 galaxies were
contributed by young, massive stars, with only a small contribution being
associated to the nuclear active source.  \citet{Cetal97} estimated that
only 1-10\% of the UV flux in 4 Seyfert~2 galaxies imaged with the HST was
originated by the nuclear source, while the rest was due to young hot stars
recently formed around the nuclei.  The contribution of star-forming
regions to the budget energy distribution of AGNs was also pointed out some
years ago by \citet{MHetal95} based in multivawelength analysis of these
type of galaxies, and by \citet{CFT95}.

Our results show that not only the UV continuum seems to be strongly
contaminated by the emission associated to starburst episodes in Seyfert~2
galaxies, but also their soft X-ray emission. This result is consistent
with the scenario assumed for Seyfert~2 galaxies, since the opaque torus
surrounding the nucleus would hide not only the UV emission associated to
the active source but also a significant part of its soft X-ray emission
due to absorption by neutral Hydrogen. For example, a detailed analysis of
the UV and X-ray emission of NGC~1068 shows that a significant fraction of 
its soft X-ray emission
could originate from the starburst episodes present around
its nucleus. On the other hand, the hard X-ray flux would be 
 underestimated by
our models by more than two orders of magnitude, indicating that it is
mostly contributed by the active source (Jim\'enez-Bail\'on et al., in
preparation). A more extended analysis of some Seyfert~2 galaxies will be
presented elsewhere.

\section{Conclusions}
\label{sec:summ}

In this work we have explored the X-ray emission originated in a
star-forming region with only single stellar populations.  The mechanical
energy injected into the interstellar medium by stellar winds and supernova
explosions will heat the diffuse gas to very high temperatures, and will be
finally reprocessed into soft X-ray emission. We have found that the X-ray
emission observed in starburst galaxies can be well explained assuming that
a moderate fraction of the mechanical energy is finally reprocessed into
X-ray emission. The rest of the mechanical energy released leads to the
expansion of the gas, creating bubbles and gas flows at galactic scales.
An interesting implication is that the age deduced from kinematical studies
will result systematically lower than the one obtained from the analysis of
the emission line spectrum if the heating of the gas and the X-ray emission
is not taken into account.  Additionally, the inclusion of X-ray transient
systems, like SN explosions and Be/X-ray binaries showing bursts of X-ray
emission, may lead to some degree of variability in the high energy
emission of starburst galaxies.

We have computed the expected intensity of the nebular \ion{He}{ii}
$\lambda$4686 \AA, including the additional ionization of the gas by the
soft X-ray emission originated in the diffuse gas.  The observational
values of the \ion{He}{ii}/H$\beta$ ratios can be reproduced by our models
assuming moderate efficiencies (about 20\%) in the reprocessing of
mechanical energies into X-ray emission. Alternatively, the mechanical
energy reprocessing could not be a continuous process. This energy released
by the massive stars could be ``accumulated'' in the form of accelerated
gas flows, which would release all this accumulated energy only when the
outflowing gas interacts with the static interstellar medium. Under this
scenario the soft X-ray emission originated by the shocked gas would
provide enough additional ionizing power to explain the observed
\ion{He}{ii}/H$\beta$ ratios.  We have also shown that such ratios have a
high intrinsic dispersion and that a deeper statistical study is necessary
to investigate the source of the nebular \ion{He}{ii} $\lambda$4686 \AA
~line.

We have compared the predicted soft X-ray emission with observational
values for a sample of star-forming and Seyfert galaxies, aiming to
disentangle the contribution of star-formation episodes to the total energy
budget of low activity galaxies. We have found that while the high energy
emission of Seyfert~1 galaxies is clearly above the predictions of
starburst models, both the UV and soft X-ray emissions of many Seyfert~2
galaxies are apparently associated mostly to the (circum-)nuclear starburst
episodes known to be present in these objects. The active source in
Seyfert~2 galaxies would dominate therefore only at harder X-rays.

\begin{acknowledgements}
  
  We want to acknowledge the referee, Onno Pols, for his valuable sugestions
  that have allowed to improve the clarity of this paper.  Useful comments
  have been provided by Mar\'{\i}a de Santos and Pedro
  Rodr\'{\i}guez-Pascual. MC wants to acknowledge Daniel Schaerer for
  useful comments about the models and evolutionary tracks, Valentina
  Luridiana for useful comments about the manuscript, Roland Diehl and
  Stephan Pl\"uschke for coments about superbubble evolution and
  Gra$\dot{\mathrm{z}}$yna Stasi\'nska for very useful comments about the
  dispersion in the emission lines. MC wants to acknowledge the {\it
    Instituto de Astrof{\'\i}sica de Andaluc{\'\i}a} for logistic support.
  We want also to thank Eduardo Fernandes Vieira and Carlos Rodrigo Blanco
  for useful comments.  This work has been supported by Spanish CICYT
  ESP-95-0389-C02-02. MC has been supported by an INTA ``Rafael Calvo
  Rod\'es'' predoctoral grant, an ESA postdoctoral grant and a MPE grant.

\end{acknowledgements}

\bibliographystyle{apj}
\end{document}